\documentclass[pra, aps, twocolumn, groupedaddress, showpacs, superscriptaddress]{revtex4-2}
\usepackage{amssymb,amsmath,amsthm,color,graphicx,times,graphicx}
\usepackage{hyperref}
\usepackage{subfigure}
\usepackage{times}
\usepackage{bbold}
\usepackage{color}

\providecommand{\openone}{\leavevmode\hbox{\small1\kern-4.3pt\normalsize1}}

\theoremstyle{plain}

\theoremstyle{definition}

\begin{document}
\title{Geometrical, topological and dynamical description of $\mathcal{N}$ interacting spin-$\mathtt{s}$ under long-range Ising model and their interplay with quantum entanglement}
\author{Brahim Amghar}\email{brahim.amghar@um5s.net.ma}\affiliation{LPHE-Modeling and Simulation, Faculty of Sciences, Mohammed V University in Rabat, Rabat, Morocco.}
\author{Abdallah Slaoui}\email{abdallah.slaoui@um5s.net.ma}\affiliation{LPHE-Modeling and Simulation, Faculty of Sciences, Mohammed V University in Rabat, Rabat, Morocco.}\affiliation{Centre of Physics and Mathematics, CPM, Faculty of Sciences, Mohammed V University in Rabat, Rabat, Morocco.}
\author{Jamal Elfakir}\affiliation{LPHE-Modeling and Simulation, Faculty of Sciences, Mohammed V University in Rabat, Rabat, Morocco.}
\author{Mohammed Daoud}\affiliation{Department of Physics, Faculty of Sciences, University Ibn Tofail, Kenitra, Morocco.}

\begin{abstract}
Comprehending the connections between the geometric, topological, and dynamical structures of integrable quantum systems with quantum phenomena exploitable in quantum information tasks, such as quantum entanglement, is a major problem in geometric information science. In this work we investigate these issues in a physical system of $\mathcal{N}$ interacting spin-$\mathtt{s}$ under the long-range Ising model. We discover the relevant dynamics, identify the corresponding quantum phase space, and we derive the associated Fubini-Study metric. Through the application of the Gauss-Bonnet theorem and the derivation of the Gaussian curvature, we proved that the dynamics occurs on a spherical topology manifold. Afterwards, we analyze the gained geometrical phase under the arbitrary and cyclic evolution processes and solve the quantum brachistochrone problem by establishing the time-optimal evolution. Moreover, by narrowing the system to a two spin-$\mathtt{s}$ system, we explore the relevant entanglement from two different perspectives. The first is geometrical in nature and involves the investigation of the interplay between the entanglement degree and the geometrical structures, such as the Fubini-Study metric, the Gaussian curvature, and the geometrical phase. The second is dynamical in nature and tackles the entanglement effect on the evolution speed and geodesic distance. Additionally, we resolve the quantum brachistochrone problem based on the entanglement degree.
\par
\vspace{0.25cm}
\textbf{Keywords:} Quantum phase space, Fubini-Study metric, Gaussian curvature, Geometrical phase, Quantum brachistochrone problem, I-concurrence.
\end{abstract}
\date{\today}

\maketitle
\section{Introduction}
One of the most outstanding aspects of modern physics is the introduction of geometric concepts describing the fundamental constituents of nature. In particular, the appearance of geometric quantum mechanics, which emerged more than four decades ago through the works of Kibble \cite{Kibble1978,Kibble1979}. This geometric approach aims to reconstruct a quantum phase space that contains all possible physical states for a specific quantum system \cite{Kibble1979,Prugovecki1982,Dias2004,2001Kus,Brody2001}. Otherwise, the geometrical formulation of quantum mechanics consists in replacing the usual Hilbert space with the concept of quantum state manifold, which, in turn possesses naturally the structure of a Kähler manifold \cite{Provost1980,Bengtsson2017,Zhang1995,Zhang1990}. In recent years, the manipulation of the geometrical features characterizing the space of quantum states has played a pivotal role in the study of the physical properties of many integrable quantum systems \cite{Nielsen2006,Li2013,Kuzmak2017,Brody2006,Deffner2017}. Some of the most notable of these properties are those associated with quantum dynamics. It is demonstrated that determining the quantum phase space naturally introduces the concept of the evolution trajectory into quantum theory \cite{Cerruti2002,Amghar2020,Amghar2021,Amghar2022}. Furthermore, the geodesic distance covered by an evolving quantum system between any two quantum states is closely related to the energy uncertainty, which is correlated to the evolution speed \cite{Anandan1990}. The shortest path feasible between any two quantum states of a spin-$1/2$ system has also been demonstrated, utilizing geometrical tools identifying the Bloch sphere ${S^2} \simeq \mathbb{C}P$ \cite{Tkachuk2011,Boscain2006}. Further, the geometrical structures are always very helpful for solving the quantum brachistochrone problem, which is related to achieving time-optimal evolution \cite{Mostafazadeh2007,Borras2007,Frydryszak2008,Wang2015}. Such evolution is essential in the construction of quantum circuits that can be used to implement quantum logic gates \cite{DiVincenzo1998,Vartiainen2004,Zhu2019}. For other additional dynamical properties investigated using these geometric approaches, we advise readers to also consult the papers \cite{Brody2015,Brody2019,Pires2015}.\par

Nowadays, the geometrization of quantum theory is the foundation of geometrical science of information, in which the quantum phenomena are addressed geometrically on the relevant quantum phase space, one can cite, for example, the quantum entanglement which is a fascinating physical resource in the tasks of quantum information theory \cite{Horodecki2009,Slaoui2019,Shaukat2020,Lewenstein2007,Amico2008}. It is shown that entanglement is closely linked to the Mannoury-Fubini-Study metric, measuring the smallest geodesic distance between an entangled state and the closest disentangled state \cite{Levay2004}. The interplay between the quantum entanglement and the manifold state curvature has been extensively investigated for $\mathcal{N}$ interacting spin-$1/2$ under the all-range Ising-model \cite{Kuzmak2018}. Further to that, the geometrical interpretation of entanglement is also discussed within the scope of Hopf fibration, being  a topological map narrowing the relevant quantum-state manifold to an another lower-dimensional manifold known as the Hopf bundle \cite{Chruscinski2006,Mosseri2007}. For more results showing the connection between quantum entanglement and geometrical features, see the following references \cite{Milman2006,Heydari2006,Slaoui2020}.\par

Some other important concept that has received a lot of attention in quantum dynamics is the geometrical phase \cite{Berry1984,Aharonov1987,Pati19951,Simon1983}, which is a noticeable intrinsic property in the evolution of quantum systems. Geometrically, it is the holonomy accumulated by the state vector during parallel transport along the evolution path \cite{Andersson2019,Demler1999}. The geometric phase is now inextricably linked to the other geometric structures that characterize the quantum phase spaces. Indeed, it is established that it can be written as the integral of the Berry-Simon connection along a cyclic evolution process, and that this connection is also related by the Fubini-Study metric thru the Bergmann kernel \cite{Samuel1988,Botero2003}. Several recent studies demonstrated the valuable role of the geometrical phase in the development of quantum information theory. Indeed, it is an advantageous tool in the implementation of quantum logic gates useful in quantum computing \cite{Zhu2002,Zhu2003}. Besides that, the experimental evidence for a conditional phase gate has been supplied both through the nuclear magnetic resonance \cite{Jones2000} and trapped ions \cite{Duan2001}. In addition, the fractional geometrical and topological phases acquired by the two-qudit systems during local unitary operations have been thoroughly investigated in relation with the entanglement \cite{Oxman2011,Khoury2014}. Other geometrical and topological phase implementations are reported in the references \cite{Oxman2018,Khoury2013,Matoso2016}.\par

The primary goal of this work is to shed light on the geometrical, topological and dynamical features of a physical system consisting of $\mathcal{N}$ interacting spin-$\mathtt{s}$ under the long-range Ising model, as well as their interplay with the quantum entanglement. It is worth noting that the ideas developed in this work were mainly motivated by the results reported by Krynytskyi and Kuzmak in \cite{Kuzmak2018, Krynytskyi2019}. In fact, by investigating the dynamics of the system, we define the relevant quantum phase space and the associated Fubini-Study metric. We compute the Gaussian curvature (G-curvature) and determine the state space topology using the Gauss-Bonnet theorem. We investigate the gained geometrical phase and solve the quantum brachistochrone problem. Eventually, we provide a comprehensive description of the geometrical and dynamical structures of two interacting spin-$\mathtt{s}$, under the Ising model, in relation to the quantum entanglement.\par

Our results are organized as follows. In Sec.\ref{Sec2}, by studying the quantum evolution of the $\mathcal{N}$ spin-$\mathtt{s}$ system under the long-range Ising-model, we identify the corresponding quantum phase space determining the Fubini-Study metric. Additionally, we discover the associated topology using the Gauss-Bonnet theorem. The geometrical phase accumulated by the system during the arbitrary and cyclic evolution processes is also investigated in Sec.\ref{sec3}. The solution of the quantum brachistochrone problem is given by examining the evolution speed as well as the corresponding geodesic distance, in Sec.\ref{sec4}. In Sec.\ref{Sec5}, we study the entanglement between two interacting spin-$\mathtt{s}$ under the Ising model from two different aspects: geometric and dynamic. The first brings to light the relation between the entanglement and derived geometrical structures, such as the Fubini-Study metric, the G-curvature and the geometrical phase, while the second discusses the entanglement effect on the evolution speed and geodesic distance. Besides that, we use the entanglement degree to solve the quantum brachistochrone problem. We provide concluding remarks in Sec.\ref{sec6}.

\section{Unitary evolution, geometry and topology of $\mathcal{N}$ spin-$\mathtt{s}$ system}\label{Sec2}

\subsection{Theoretical model and unitary quantum evolution}
In this work, the considered system consists of $\mathcal{N}$ qudits (with $d=2\mathtt{s}+1$) represented by $\mathcal{N}$ interacting spin-$\mathtt{s}$ under long-range Ising model governed by the Hamiltonian
\begin{equation}\label{a}
\mathrm{H}=2\mathtt{J}\sum\limits_{1 \le k <l\le \mathcal{N}} {\mathtt{S}_k^z\mathtt{S}_l^z},
\end{equation}
with $\mathtt{J}$ is the exchange constant of the interaction and $\mathtt{S}_k^z$ stands for the $z$ component of the spin operator $\textbf{S}_k=(\mathtt{S}_k^x,\mathtt{S}_k^y,\mathtt{S}_k^z)^T$ associated with $kth$ spin-$\mathtt{s}$ (i.e., the $k$th qudit) which fulfills the eigenvalues equation of the form
\begin{equation}
\mathtt{S}_k^z\left| {{\mathsf{m}_k}} \right\rangle  = {\mathsf{m}_k}\left| {{\mathsf{m}_k}} \right\rangle,
\end{equation}
with $\mathsf{m}_{k}=\left\lbrace-\mathtt{s},-\mathtt{s} + 1,\cdots,\mathtt{s}\right\rbrace$ are the possible values due to the projection of the $k$th spin over the $z$ axis, and $\left| {{\mathsf{m}_k}} \right\rangle $ represent the associated eigenstates. It is worth noting that the components of spin-$\mathtt{s}$ operators $\mathtt{S}_k^x,\,\mathtt{S}_k^y,$ and $\mathtt{S}_k^z$ fulfill the algebraic structure of the Lie group SU(2),
\begin{equation}
\left[ {\mathtt{S}_k^\alpha ,\mathtt{S}_l^\beta } \right] = i{\delta _{kl}}\sum\limits_{\gamma  = x,y,z}\epsilon^{\alpha\beta\gamma} {\mathtt{S}_k^\gamma },
\end{equation}
where $\delta _{kl}$ and $\epsilon^{\alpha\beta\gamma}$ denote the Kronecker and Levi-Civita symbols, respectively. We presume that the system is initially maintained in a coherent state achieved by a rotation of the maximum weight state $\left|{\mathtt{s}, \mathtt{s},\cdots,\mathtt{s}} \right\rangle$ (i.e., all the spins take its maximal values) through an angle $\Theta$ around the axis $\textbf{r}=(\sin\Phi,-\cos\Phi,0)$. It is then explicitly formulated by
\begin{equation}\label{b}
\left| {{\Psi _i}} \right\rangle  = {e^{ - i\Theta \sum\limits_{k = 1}^\mathcal{N} {{\textbf{S}_k}.\textbf{r}} }}\left| { \mathtt{s}, \mathtt{s},\cdots,\mathtt{s}} \right\rangle = {e^{ \sum\limits_{k = 1}^\mathcal{N} {({\mu} \mathtt{S}_k^ +  - {\mu}^* \mathtt{S}_k^ - )} }}\left| {  \mathtt{s},  \mathtt{s},\cdots,\mathtt{s}} \right\rangle,
\end{equation}
where ${\mu}=\frac{\Theta }{2}{e^{-i\Phi}}$, such as $\Theta\in[0,\pi]$ and $\Phi\in[0,2\pi]$ represent, respectively, the polar and azimuthal angles. The initial state \eqref{b} can be also written as \cite{Zhang1990}
\begin{equation}\label{d}
\left| {{\Psi _i}} \right\rangle  = {(1 + \mathtt{Z}{\mathtt{Z}^*})^{ - \mathcal{N}\mathtt{s}}}\left\| {\left. {{\mathtt{Z}_1},{\mathtt{Z}_2},\cdots,{\mathtt{Z}_\mathcal{N}}} \right\rangle } \right.,
\end{equation}
with
\begin{equation}
\left\| {\left. {{\mathtt{Z}_k}} \right\rangle } \right. = \sum\limits_{{\mathsf{m}_k} =  - \mathtt{s}}^\mathtt{s} {{\mathtt{Z}^{\mathtt{s} + {\mathsf{m}_k}}}\sqrt {C_{2\mathtt{s}}^{\mathtt{s}+\mathsf{m}_k}} } \left| {{\mathsf{m}_k}} \right\rangle,
\end{equation}
is the non-normalized SU(2) coherent state for the $k$th qudit, $C$ stands for the binomial coefficient, and the complex parameter $\mathtt{Z}= \tan \frac{\Theta }{2}{e^{ - i\Phi }}$ defines the stereographic projection of the coherent state space (sphere) $S^2\simeq\mathbb{C}P\simeq{{\rm SU(2)} \mathord{\left/
 {\vphantom {{\rm SU(2)} {\rm U(1)}}} \right.
 \kern-\nulldelimiterspace} {\rm U(1)}}$ from the north pole on the relevant equatorial plane (i.e., local coordinate on $\mathbb{C}P$). Further, the state-space geometry resulting from the rotation of the maximum weight state can be determined using the Fubini-Study metric which is defined by \cite{Zhang1995,Bengtsson2017}
 \begin{equation}\label{am1}
dS^2 = \frac{{{\partial ^2}\ln{\mathsf{K}(\mathtt{Z},{\mathtt{Z}^*})}}}{{\partial \mathtt{Z}\partial {\mathtt{Z}^*}}}d\mathtt{Z}d{\mathtt{Z}^*},
\end{equation}
with 
\begin{equation}
\mathsf{K}(\mathtt{Z},\mathtt{Z}^*)=\left\langle {{\mathtt{Z}_1},{\mathtt{Z}_2},\cdots,{\mathtt{Z}_\mathcal{N}}} \right.\left\| {\left. {{\mathtt{Z}_1},{\mathtt{Z}_2},\cdots,{\mathtt{Z}_N}} \right\rangle } \right.=(1+\mathtt{Z}\mathtt{Z}^*)^{2\mathcal{N}\mathtt{s}},
\end{equation}
is the Bergmann kernel characterizing the resulting state manifold. By a straightforward calculation, the Fubini-Study metric \eqref{am1} can be written as
\begin{equation}\label{f}
dS^2 = \frac{{\mathcal{N}\mathtt{s}}}{2}\left[d{\Theta ^2} + {\sin ^2}\Theta d{\Phi ^2}\right].
\end{equation}
It follows that the space of initial state \eqref{d} is indeed a sphere of radius $\sqrt {{{\mathcal{N}\mathtt{s}} \mathord{\left/
 {\vphantom {{N\mathtt{s}} 2}} \right.
 \kern-\nulldelimiterspace} 2}}$. We will now evolve the initial state \eqref{d} via the time evolution propagator $\mathcal{P}(t)=e^{-i\mathrm{H}t}$. As a result, the evolved state of the $\mathcal{N}$ spin-$\mathtt{s}$ system is found as
 \begin{widetext}
 \begin{equation}\label{e}
\left|\Psi \left( t\right)\right\rangle= {(1 + \mathtt{Z}{\mathtt{Z}^*})^{ - \mathcal{N}\mathtt{s}}}\sum\limits_{{\mathsf{m}_1},{\mathsf{m}_2},...,{\mathsf{m}_\mathcal{N}} =  - \mathtt{s}}^\mathtt{s} {{e^{ - i2\xi(t) \sum\nolimits_{1 \le k < l \le \mathcal{N}} {{\mathsf{m}_k}{\mathsf{m}_l}} }}\left( {\prod\limits_{\nu  = 1}^\mathcal{N} {{\mathtt{Z}^{\mathtt{s} + {\mathsf{m}_\nu  }}}} \sqrt {C_{2\mathtt{s}}^{\mathtt{s} + {\mathsf{m}_\nu  }}} } \right)} \left| {{\mathsf{m}_1},{\mathsf{m}_2},\cdots,{\mathsf{m}_\mathcal{N}}} \right\rangle,
\end{equation}
\end{widetext}
where $\xi(t)=\mathtt{J}t$. It should be noted that the evolution of the system is ensured only by the parameter $\xi$, while the other parameters, namely, $\mathcal{N},\mathtt{s},\Theta$, and $\Phi$ specify the choice of the initial state retained. Also, it is interesting to note that the $\mathcal{N}$ spin-$\mathtt{s}$ state \eqref{e} fulfills the following periodic requirements:
\begin{equation}
\left| \Psi(\xi+2\pi) \right\rangle =\left| \Psi(\xi) \right\rangle\quad\quad \text{if}\; \mathtt{s}\; \text{half-integer},
\end{equation}
and
\begin{equation}
\left| \Psi(\xi+\pi) \right\rangle =\pm\left| \Psi(\xi) \right\rangle\quad\quad \text{if}\; \mathtt{s}\; \text{integer}.
\end{equation}
Thus, the behavior of the wave function describing the collection of $\mathcal{N}$ particles is then periodic along the parameter $\xi$ with a period depending on the bosonic (integer spin) or fermionic (half-integer spin) character of the particles under study.
\subsection{Geometry and topology of the resulting state space}
After performing the evolution of the $\mathcal{N}$ particle system via the temps-evolution propagator and identifying the evolved state \eqref{e}, let us now explore the geometry and the topology of the resulting quantum state manifold containing all the states that the system can attain during the evolution. For this purpose, we employ the Fubini-Study metric which is defined by the infinitesimal distance $d{\mathtt{\bf{S}}}$ between two neighboring pure quantum states $|\Psi(\zeta^a)\rangle$ and $|\Psi(\zeta^a+\mathrm{d} \zeta^a)\rangle$ and given by \cite{Abe1993,Tkachuk2011}
\begin{equation}\label{am2}
d{\mathtt{\bf{S}}}^2= {\mathrm{g}_{ab}}d{\zeta ^a}d{\zeta ^b},
\end{equation}
where $\zeta^a$ are the freedom degrees $\Theta$, $\Phi$, and $\xi$ defining the evolving state \eqref{e} and $\mathrm{g}_{ab}$ represent the components of this metric having the form
\begin{equation}
\mathrm{g}_{a b}=\mathrm{Re}\left(\left\langle\Psi_{a}| \Psi_{b}\right\rangle-\left\langle\Psi_{a} | \Psi\right\rangle\left\langle\Psi|\Psi_{b}\right\rangle\right),
\end{equation}
with $\left| {{\Psi _{a,b}}} \right\rangle  = \frac{\partial }{{\partial {\zeta ^{a,b}}}}\left| \Psi  \right\rangle $. Using the definition \eqref{am2} and taking into account the binomial theorem
\begin{equation}\label{ppm}
\sum\limits_{{\mathsf{m}_k} =  - \mathtt{s}}^\mathtt{s} {{{\left( {{{\tan }^2}\frac{\Theta }{2}} \right)}^{\mathtt{s} + {\mathsf{m}_k}}}} C_{2\mathtt{s}}^{\mathtt{s} + {\mathsf{m}_k}} = {\left( {1 + {{\tan }^2}\frac{\Theta }{2}} \right)^{2\mathtt{s}}},
\end{equation}
we get the explicit form  of the metric tensor \eqref{am2} as
		\begin{small}
			\begin{align}\label{g}
			d{\mathtt{\bf{S}}}^2=&dS^2+  \frac{1}{2}\mathcal{N}(\mathcal{N} - 1){\mathtt{s}^2}{\sin ^2}\Theta \left[ {1 + (4\mathtt{s}(\mathcal{N} - 1) - 1){{\cos }^2}\Theta } \right]d\xi^2 \notag\\&+ \mathcal{N}(\mathcal{N} - 1){\mathtt{s}^2}\cos \Theta {\sin ^2}\Theta d{\Phi} d{\xi},
			\end{align}
		\end{small}
where $dS^2$ is the line element given in \eqref{f}. Thus, we managed to establish the Riemannian geometry of the resulting state manifold after performing the temporal evolution of the $\mathcal{N}$ spin-$\mathtt{s}$ system. It is simple to verify that for $\xi=0$ (no evolution), the space of $\mathcal{N}$ spin-$\mathtt{s}$ states \eqref{g} reduces to the sphere \eqref{f}. As we can see, the components of the metric tensor \eqref{g} are $\Phi$ independent, meaning that the quantum state spaces with a predetermined azimuthal angle possess the same geometry. Accordingly, we conclude that the space of states (i.e., quantum phase space) associated with the $\mathcal{N}$ qudits (under study) is a curved two-dimensional manifold being parameterized by $\Theta$ and $\xi$. Therefore, it is defined by the following metric tensor
\begin{small}
\begin{align}\label{m}
d{\mathtt{\bf{S}}}^{2}=&\frac{1}{2}\mathcal{N}(\mathcal{N} - 1){\mathtt{s}^2}{\sin ^2}\Theta \left\lbrace {1 + \left[4\mathtt{s}(\mathcal{N} - 1) - 1\right]{{\cos }^2}\Theta }\right\rbrace d\xi^{2}\notag\\&+\frac{{\mathcal{N}\mathtt{s}}}{2}d{\Theta ^2}.
\end{align}
\end{small}
Here, we are interested in the topological aspect of the $\mathcal{N}$ spin-$\mathtt{s}$ system by determining the topology of the relevant quantum state manifold. To realize this, we start by computing G-curvature which can be expressed in terms of the metric tensor \eqref{m} as follows \cite{Kolodrubetz2013}:
\begin{align}
\mathit{K} = \frac{1}{{{{({\mathrm{g}_{_{\Theta \Theta }}}{\mathrm{g}_{_{\xi \xi} }})}^{1/2}}}}&\left[ \frac{\partial }{{\partial \xi }}\left( {{{\left( {\frac{{{\mathrm{g}_{\xi \xi }}}}{{{\mathrm{g}_{_{\Theta \Theta} }}}}} \right)}^{1/2}}\Gamma _{_{\Theta \Theta }}^\xi } \right) \right.\notag\\&- \left.\frac{\partial }{{\partial \Theta }}\left( {{{\left( {\frac{{{\mathrm{g}_{_{\xi \xi }}}}}{{{\mathrm{g}_{_{\Theta \Theta} }}}}} \right)}^{1/2}}\Gamma _{_{\Theta \xi }}^\xi } \right) \right],
\label{h}	
\end{align}
with $\Gamma_{\Theta \Theta}^{\xi}$ and $\Gamma_{\Theta \xi}^{\xi}$ are the Christoffel symbols take the form 
\begin{equation}
\Gamma_{\Theta \Theta}^{\xi}=-\frac{1}{2 \mathrm{g}_{_{\xi \xi}}}\left(\frac{\partial \mathrm{g}_{_{\Theta \Theta}}}{\partial \xi}\right),\hspace{0.4cm}{\rm and}\hspace{0.4cm}\Gamma_{\Theta \xi}^{\xi}=\frac{1}{2 \mathrm{g}_{_{\xi \xi}}}\left(\frac{\partial \mathrm{g}_{_{\xi \xi}}}{\partial \Theta}\right).
\end{equation}
Notice that the metric component $\mathrm{g}_{_{\xi\xi}}$  vanishes at the points $\Theta=0$ or $\Theta=\pi$, hence the G-curvature \eqref{h} of the quantum state manifold \eqref{m} is not defined in these two points. Thus, we can say that the G-curvature has two singularities, one in $\Theta=0$ and the other in $\Theta=\pi$. Otherwise, the curvature is defined in all other points of the manifold. Inserting the metric components in the equation \eqref{h}, we obtain the G-curvature written as
\begin{equation}\label{ay}
\mathit{K} = \frac{4}{{\mathcal{N}\mathtt{s}}}\left( 2 - \frac{\left\lbrace {4\mathtt{s}(\mathcal{N} - 1) -1} \right\rbrace{{\cos }^2}\Theta  + 2\mathtt{s}(\mathcal{N} - 1) + 1}{{{{\left[\left\lbrace 4\mathtt{s}(\mathcal{N} - 1) - 1 \right\rbrace {{\cos }^2}\Theta +1 \right]}^2}}} \right).
\end{equation}
Interestingly, the G-curvature \eqref{ay} of the resulting quantum phase space \eqref{m} depends on the initial parameters $(\Theta,\, \mathcal{N}, \,\mathtt{s})$ of the considered system, while it is independent of the time, which clearly shows that the state-space geometry is not affected by the dynamics of the system. On the other hand, it is straightforward to check that for $\mathcal{N}>2$ and $\mathtt{s} \ge \frac{1}{2}$, the G-curvature can take negative values for some values of $\Theta$. This is consistent with the outcomes reported in \cite{Krynytskyi2019}.\par
Given this fact and the fact that the G-curvature \eqref{ay} possesses two singularities, we conclude that the relevant quantum state manifold \eqref{m} includes two conical defects: the first one is located close to the point $\Theta=0$, while the second one is located close to the point $\Theta=\pi$. Let us now determine the topology associated with the space of $\mathcal{N}$ spin-$\mathtt{s}$ states \eqref{m}. For this, we need to calculate the integer Euler characteristic $\chi(\mathrm{M})$ [$\mathrm{M}$ represents the state manifold \eqref{m}] given in the Gauss-Bonnet theorem as follows \cite{Kolodrubetz2013}
\begin{equation}\label{am44}
\frac{1}{2 \pi}\left[\int_{\mathrm{M}} \mathit{K} d \mathtt{S}+\oint_{\partial \mathrm{M}} k_{g} d l\right]=\chi(\mathrm{M})
\end{equation}
where $d\mathtt{S}$ and $dl$ are, respectively, the area and line elements on $\mathrm{M}$, while $k_g$ stands for the geodesic curvature. Further, the first and second terms on the left side of equation \eqref{am44} indicates, respectively, the bulk and boundary contributions to the Euler characteristic specifying the state manifold topology. We can easily verify that the Gauss-Bonnet theorem \eqref{am44} can be written in terms of the Fubini-Study metric \eqref{m} as  
\begin{equation}\label{i}
\int_0^{\pi} {\int_0^{{\xi _{\max }}} {\textit{K}{{\left( {{\mathrm{g}_{_{\Theta \Theta} }}{\mathrm{g}_{_{\xi \xi} }}} \right)}^{1/2}}} } d\Theta d\xi  + \Delta  = 2\pi\chi(\mathrm{M}),
\end{equation}
with $\Delta$ corresponds to the Euler boundary integral containing the conical defects contribution. Explicitly, the Gauss-Bonnet theorem \eqref{i} rewrites
\begin{equation}\label{aml2}
4\mathtt{s}{\xi _{\max }}(\mathcal{N} - 1)+ \Delta= 2\pi\chi(\mathrm{M}).
\end{equation}
Thus, to compute the Euler characteristic $\chi$, we must first determine the Euler boundary integral $\Delta$. For this aim, one assumes that the angular defects are located very close to the singular points $\Theta=0,\pi$. In this respect, the metric tensor \eqref{m} can be expanded, close to these two singular points, up to the second order in $\Theta$. Indeed, we find
\begin{equation}\label{k}
d{\mathtt{\bf{S}}}^2=\frac{{\mathcal{N}\mathtt{s}}}{2}d{\Theta ^2}+ 2\mathcal{N}(\mathcal{N}-1)^2\mathtt{s}^3\Theta^2d\xi^2.
\end{equation}
In addition, it is worth noting that the solid angle of a revolution cone of angle at the apex $2\Theta$ is defined by $\Omega=2\pi\left(1-\cos\Theta\right)$ where the last term of this expression stands for the partial solid angle traced out by the system around the cone apex in the evolution. Taking advantage of the proximity of the angular defects to the two singular points, we simply find
\begin{equation}
2\pi\cos\Theta\approx \frac{{{{\mathtt{\bf{S}}}}({\xi _{\max }})}}{\mathbf{d}}=\dfrac{\sqrt{\mathrm{g}_{_{\xi\xi}}}\xi_{\max}}{\sqrt{\mathrm{g}_{_{\Theta\Theta}}}\Theta},
\end{equation}
with ${\mathtt{\bf{S}}}(\xi_{\max})$ representing the distance covered by the system in the time period $\textbf{t} = {{{\xi _{\max }}} \mathord{\left/
 {\vphantom {{{\xi _{\max }}} \mathtt{J}}} \right.
 \kern-\nulldelimiterspace} \mathtt{J}}$ around one of the two singular points ($\Theta=0$ or $\pi$) and $\mathbf{d}$ denoting the distance between the trajectory of the system and the relevant singular point. Consequently, we obtain the angular defects as follows
 \begin{equation}\label{l}
\Delta=2\left[2\pi- \frac{{\sqrt {{\mathrm{g}_{_{\xi \xi} }}} {\xi _{\max}}}}{{\sqrt {{\mathrm{g}_{_{\Theta \Theta} }}} \Theta }}\right]=2\left[2\pi-2\mathtt{s}\xi_{\max}(\mathcal{N}-1)\right],
\end{equation}
where we multiplied it by the factor of $2$ since we have two singular points. Reporting this last equation into the Gauss-Bonnet formula \eqref{aml2}, we obtain the integer Euler characteristic  $\chi(\mathrm{M})= 2$, meaning that the quantum state space \eqref{m} of the $\mathcal{N}$ spin-$\mathtt{s}$ system has the topology of a sphere. The upcoming section will reserved for the study of the geometric phases that can be accumulated by the $\mathcal{N}$ spin-$\mathtt{s}$ during the evolutionary process on the resulting state manifold.
\section{Geometric phases accumulated by the $\mathcal{N}$ spin-$\mathtt{s}$ system}\label{sec3}
Having exploring the geometry and the topology of the quantum state space specified by the squared line element \eqref{m}. Let us now examine the geometric phase that can be acquired by the evolving state \eqref{e} for arbitrary and cyclic evolution processes.
\subsection{Geometric phase in an arbitrary evolution}
At this stage, we consider that the $\mathcal{N}$ spin-$\mathtt{s}$ system evolves arbitrarily on the two-dimensional manifold \eqref{m}. In this scheme, the geometrical phase acquired by the evolved state \eqref{e} has the following form \cite{Mukunda1993,Oxman2011}:
		\begin{equation}\label{y}
		{\boldsymbol{\Phi}_\text{g}}(t) = \arg \langle {\Psi _i}|{\Psi (t)}\rangle - {\mathrm{Im}}\int_0^t {\left\langle {\Psi (t')} \right|} \frac{\partial }{{\partial t'}}\left| {\Psi (t')} \right\rangle dt^\prime,
		\end{equation}
being, of course, the global phase minus the dynamical phase. To evaluate the geometrical phase, we first derive the global phase accumulated by the system. The transition-probability amplitude (i.e., the overlap) between the initial state \eqref{d} and the evolved state \eqref{e} can be written as
\begin{widetext}
\begin{align}\label{aw}
	\left\langle\Psi _{i} \mid \Psi \left(t \right) \right\rangle = {\left( {1 + {{\tan }^2}\frac{\Theta }{2}} \right)^{ - 2\mathcal{N}\mathtt{s}}}\sum\limits_{{\mathsf{m}_1},...,{m_\mathcal{N}} =  - \mathtt{s}}^\mathtt{s} {{e^{ - i2\xi \sum\nolimits_{1 \le k < l \le \mathcal{N}} {{\mathsf{m}_k}{\mathsf{m}_l}} }}\left[ {\prod\limits_{\nu   = 1}^\mathcal{N} {{{\left( {\tan \frac{\Theta }{2}} \right)}^{2(\mathtt{s} + {\mathsf{m}_\nu  })}}} C_{2\mathtt{s}}^{\mathtt{s} + {\mathsf{m}_\nu  }}} \right]}. 
\end{align}
Reporting this last equation \eqref{aw} into the first term on the right-hand side in equation \eqref{y}, the total phase accumulated by the $\mathcal{N}$ spin-$\mathtt{s}$ state reads as
\begin{equation}\label{ab}
{\boldsymbol{\Phi}_{\text{glob}}} =  - \arctan \left( {\frac{{\sum\limits_{{\mathsf{m}_1},{\mathsf{m}_2},...,{\mathsf{m}_\mathcal{N}} =  - \mathtt{s}}^\mathtt{s} {\sin \left( {2\xi \sum\limits_{1 \le k < l \le \mathcal{N}} {{\mathsf{m}_k}{\mathsf{m}_l}} } \right)\left( {\prod\limits_{\nu   = 1}^\mathcal{N} {{{\left( {\tan \frac{\Theta }{2}} \right)}^{2(\mathtt{s} + {\mathsf{m}_\nu  })}}C_{2s}^{\mathtt{s} + {\mathsf{m}_\nu  }}} } \right)} }}{{\sum\limits_{{\mathsf{m}_1},{\mathsf{m}_2},...,{\mathsf{m}_\mathcal{N}} =  - \mathtt{s}}^\mathtt{s} {\cos \left( {2\xi \sum\limits_{1 \le k < l \le \mathcal{N}} {{\mathsf{m}_k}{\mathsf{m}_l}} } \right)\left( {\prod\limits_{\nu   = 1}^\mathcal{N} {{{\left( {\tan \frac{\Theta }{2}} \right)}^{2(\mathtt{s} + {\mathsf{m}_\nu  })}}C_{2\mathtt{s}}^{\mathtt{s} + {\mathsf{m}_\nu  }}} } \right)} }}} \right).
\end{equation}
\end{widetext}
It seems that the global phase is composed of two main parts: the first part is of geometrical nature (i.e., the geometrical phase) being profoundly linked to the geometrical and topological features of these systems. This part can be explained by the explicit dependence of the global phase upon the dynamical parameters $(\Theta,\xi)$ defining the evolution path on the quantum phase space \eqref{m}, which means that this phase depends both on the evolution path and the geometry (or the topology) in the state manifold \eqref{m}. The second part is of dynamical nature (i.e., the usual dynamical phase) which can be justified by the dependence of the global phase on the average value of the Hamiltonian \eqref{a}. Moreover, it is simple to check that the global phase is defined modulo 2$\pi$ and verifies the periodic requirement
\begin{equation}
{\boldsymbol{\Phi}_{\text{glob}}}(\xi  + 2\pi ) = {\boldsymbol{\Phi} _{\text{glob}}}(\xi ).
\end{equation}
On the other hand, the dynamical phase can be derived by inserting the evolving state \eqref{e} into the second term on the right-hand side of equation \eqref{y}. Explicitly, we obtain
\begin{equation}\label{ac}
{\boldsymbol{\Phi} _{\text{dyn}}} =  - \xi {\mathtt{s}^2} \mathcal{N}(\mathcal{N} - 1){\cos ^2}\Theta.
\end{equation}
Hence, the geometrical phase accumulated by the $\mathcal{N}$ spin-$\mathtt{s}$ system \eqref{e}, undergoing an arbitrary evolution over the quantum state space \eqref{m}, is then given by
\begin{widetext}
\begin{equation}\label{z}
{\boldsymbol{\Phi}_\text{g}} =  - \arctan \left( {\frac{{\sum\limits_{{\mathsf{m}_1},{\mathsf{m}_2},...,{\mathsf{m}_\mathcal{N}} =  - \mathtt{s}}^\mathtt{s} {\sin \left( {2\xi \sum\limits_{1 \le k < l \le \mathcal{N}} {{\mathsf{m}_k}{\mathsf{m}_l}} } \right)\left( {\prod\limits_{\nu   = 1}^\mathcal{N} {{{\left( {\tan \frac{\Theta }{2}} \right)}^{2(\mathtt{s} + {\mathsf{m}_\nu  })}}C_{2\mathtt{s}}^{\mathtt{s} + {\mathsf{m}_\nu  }}} } \right)} }}{{\sum\limits_{{\mathsf{m}_1},{\mathsf{m}_2},...,{\mathsf{m}_\mathcal{N}} =  - \mathtt{s}}^\mathtt{s} {\cos \left( {2\xi \sum\limits_{1 \le k < l \le \mathcal{N}} {{\mathsf{m}_k}{\mathsf{m}_l}} } \right)\left( {\prod\limits_{\nu   = 1}^N {{{\left( {\tan \frac{\Theta }{2}} \right)}^{2(\mathtt{s} + {\mathsf{m}_\nu  })}}C_{2\mathtt{s}}^{\mathtt{s} + {\mathsf{m}_\nu  }}} } \right)} }}} \right) + \xi \mathcal{N}(\mathcal{N} - 1){\mathtt{s}^2}{\cos ^2}\Theta .
\end{equation}
\end{widetext}
Unlike the dynamical phase, it is clear that the resulting geometrical phase \eqref{z} evolves non-linearly with the time. This phase depends on the dynamical parameters $(\Theta, \xi)$, meaning that it depends on the form of the evolution trajectory followed by the $\mathcal{N}$ spin-$\mathtt{s}$ state \eqref{e} over the quantum phase space \eqref{m}, whereas the dependence on the initial parameters $(\mathcal{N},\mathtt{s})$ can be interpreted by the dependence of the geometrical phase on the state space geometry. Let us now consider a special case in which we investigate the geometrical phase accumulated by the $\mathcal{N}$ spin-$\mathtt{s}$ state \eqref{e} during a very short time period. In this respect, by expanding the exponential factor in \eqref{aw} upto the second order in $\xi$, we find
\begin{widetext}
\begin{equation}\label{poste}
\left\langle {{\Psi _i}} \right|\left. {\Psi (t)} \right\rangle  \simeq 1 - \frac{{{\xi ^2}{\mathtt{s}^2}\mathcal{N}(\mathcal{N} - 1)}}{4}\left[ {\mathtt{s}(\mathcal{N} - 1)\left( {2\mathtt{s}\mathcal{N}{{\cos }^4}\Theta  + {{\sin }^2}2\Theta } \right) + {{\sin }^4}\Theta } \right] - i\xi {\mathtt{s}^2}\mathcal{N}(\mathcal{N} - 1){\cos ^2}\Theta.
\end{equation}
In this view, the geometrical phase \eqref{z} reads as
\begin{equation}\label{an}
{\boldsymbol{\Phi}_\text{g}}  \simeq   - \arctan \left( {\frac{{4\xi {\mathtt{s}^2}\mathcal{N}(\mathcal{N} - 1){{\cos }^2}\Theta }}{{4 - {\xi ^2}{\mathtt{s}^2}\mathcal{N}(\mathcal{N} - 1)\left( {\mathtt{s}(\mathcal{N} - 1)(2\mathtt{s}\mathcal{N} {{\cos }^4}\Theta  + {{\sin }^2}2\Theta ) + {{\sin }^4}\Theta } \right)}}} \right)+\xi {\mathtt{s}^2} \mathcal{N}(\mathcal{N} - 1){\cos^2}\Theta.
\end{equation}
\end{widetext}
It is interesting to note that for $\xi=0$ the system does not acquire any phase, since the evolved state \eqref{e} coincides with the initial state \eqref{d}. Moreover, we can observe that, in the thermodynamic limit $(\mathcal{N} \to \infty)$ the global phase disappears. This implies that the geometric phase and the dynamical phase are equal throughout the evolution process of the system, which gives us the possibility to measure experimentally the geometric phase since it can be expressed in terms of the Hamiltonian \eqref{a}. A similar outcome can be obtained for particles with large spin values $(\mathtt{s} \to \infty)$.
\subsection{Geometrical phase in a cyclic evolution}
Here, we focus on investigating the geometrical phase emerging from the cyclic evolution of the $\mathcal{N}$ spin-$\mathtt{s}$ system. In this scheme, the evolving state \eqref{e} fulfills the cyclic condition $\left| {\Psi (\textbf{T})} \right\rangle= {e^{i{\boldsymbol{\Phi} _{\text{glob}}}}}\left| {\Psi (0)} \right\rangle$ where $\textbf{T}$ denotes the required time for a cyclic evolution process. The AA-geometrical phase (Aharonov–Anandan phase) accumulated by the system during a cyclic evolution (i.e., enclosed path in the relevant parameter space) reads as \cite{Aharonov1987,Pati1995}
\begin{equation}\label{aa}
 \boldsymbol{\Phi}_\text{g}^{\text{AA}} = i\int_0^\textbf{T} {\left\langle {\tilde \Psi (t)} \right|} \frac{\partial }{{\partial t}}\left| {\tilde \Psi (t)} \right\rangle dt,
\end{equation}
where $\left| {\tilde \Psi (t)} \right\rangle$ standing for  the cross section introduced by Anandan and Aharonov in Ref.\cite{Aharonov1987}. It is  given in terms of the evolving state \eqref{e} as $
\left| {\tilde \Psi (t)} \right\rangle  = {e^{ - i{\mathrm{f}}(t)}}\left| {\Psi (t)} \right\rangle $ where $\mathrm{f}(t) $ is any smooth function verifying ${\mathrm{f}}(\textbf{T})-{\mathrm{f}}(0)=\boldsymbol{\Phi}_{\text{glob}}$. Therefore, the AA-geometrical phase \eqref{aa} can be rewritten as
\begin{equation}\label{ad}
\boldsymbol{\Phi}_\text{g}^{\text{AA}}=\int_0^\textbf{T} {d{\boldsymbol{\Phi} _{\text{glob}}} + i} \int_0^\textbf{T} {\left\langle {\Psi (t)} \right|} \frac{\partial }{{\partial t}}\left| {\Psi (t)} \right\rangle dt.
\end{equation}
Setting equations \eqref{ab} and \eqref{ac} into \eqref{ad}, the AA-geometric phase accumulated by the $\mathcal{N}$ spin-$\mathtt{s}$ state \eqref{e}, experiencing any cyclic evolution over the quantum state manifold \eqref{m}, is of the form
\begin{equation}
\boldsymbol{\Phi}_\text{g}^{\text{AA}}={\xi _{\max }}\mathcal{N}(\mathcal{N} - 1){\mathtt{s}^2}{\cos ^2}\Theta.
\end{equation}
Remark that the integral of the global phase vanishes $(\boldsymbol{\Phi}_{\text{glob}}=0)$ due to the cyclic evolution of the $\mathcal{N}$ spin-$\mathtt{s}$ system. Consequently, the AA-geometrical phase is the cyclic integral, in the interval $[0, \xi_{\max}]$, of the average value of the Hamiltonian \eqref{a}. Otherwise, during the parallel transport the state vector \eqref{e} over the quantum phase space \eqref{m}, the AA-geometrical phase gained coincides with the dynamical phase. Thus, the calculation of the AA-phase geometrical requires the knowledge of the Ising Hamiltonian \eqref{a}, 
which reflects the possibility of measuring it experimentally. This can be related to the time-optimal evolution of the considered system by examining the evolution speed and the corresponding geodesic distance. This is the quantum brachistochrone problem that we will solve in the next section.
\section{Evolution speed and Quantum brachistochrone problem for $\mathcal{N}$ spin-$\mathtt{s}$ system}\label{sec4}
The quantum brachistochrone problem consists essentially in establishing the optimal time evolution of such a system \cite{Mostafazadeh2007,Borras2007,Amghar2020}. The solution of this problem involves finding the smallest possible duration such that the system travels the smallest path between the departure state \eqref{d} and the arrival state \eqref{e}. The main objective in this section is to achieve the time-optimal evolution by maximizing the evolution speed and minimizing the corresponding geodesic distance carried by the system.
\subsection{Evolution speed of the system}
 To compute the evolution speed, we presume that the evolution of $\mathcal{N}$ spin-$\mathtt{s}$ system is only dependent on time, whereas all other parameters are constant. In this respect, the Fubini-Study metric \eqref{m} reduces to
 \begin{small}
 \begin{equation}\label{o}
d{\mathtt{\bf{S}}}^2=\frac{1}{2}\mathcal{N}(\mathcal{N} - 1){\mathtt{s}^2}{\sin ^2}\Theta \left\lbrace 1 + \left[4\mathtt{s}(\mathcal{N} - 1) - 1\right]{{\cos }^2}\Theta \right\rbrace d\xi^2.
\end{equation}
 \end{small}
It results that the evolution of the system happens over a circle of radius $\sqrt{g_{\xi\xi}}$ for half-integer $\mathtt{s}$ and ${{\sqrt {{g_{\xi \xi }}} } \mathord{\left/
 {\vphantom {{\sqrt {{g_{\xi \xi }}} } 2}} \right.
 \kern-\nulldelimiterspace} 2}$ for integer $\mathtt{s}$. The evolution speed of $\mathcal{N}$ spin-$\mathtt{s}$ system  reads as \cite{Anandan1990}
\begin{equation} \label{n}
\mathcal{V}=\frac{d{\mathtt{\bf{S}}}}{dt}=\dfrac{2}{\hbar}\Delta E,
\end{equation}
with $\Delta E$ is the energy uncertainty associated with the Ising Hamiltonian \eqref{a}. From the equation \eqref{n}, we can observe that the greater the energy uncertainty, the shorter the evolution time, and vice versa. By inserting equation \eqref{o} into equation \eqref{n}, the evolution speed of the $\mathcal{N}$ spin-$\mathtt{s}$ system is given by
\begin{equation}\label{p}
\mathcal{V} = \mathtt{J}\mathtt{s}\sqrt {\frac{{\mathcal{N}(\mathcal{N} - 1){{\sin }^2}\Theta \left\lbrace {1 + \left[4\mathtt{s}(\mathcal{N} - 1) - 1\right]{{\cos }^2}\Theta } \right\rbrace}}{2}}.
\end{equation}
The evolution speed is affected both by the exchange interaction between spins $\mathtt{J}$, the particle number $\mathcal{N}$, and the spin value $\mathtt{s}$. Furthermore, the greater these physical parameters are, the faster the evolution of the system is, except for $\Theta=0$ or $\Theta=\pi$, where the evolution speed vanishes $(\mathcal{V}=0)$ (no evolution) whatever these physical parameters are. This is due to the fact that the $\mathcal{N}$ spin-$\mathtt{s}$ state is not defined at these singular points. 
\subsection{Quantum brachistochrone problem}
To derive the smallest possible duration needed to realize the time-optimal evolution, we start by maximizing the evolution speed \eqref{p} by solving the equation ${{d\mathcal{V}}}/{{d\Theta}}=0$. As a result, one finds
\begin{equation}\label{r}
\sin 2\Theta \left\lbrace {2\mathtt{s}(\mathcal{N}-1)-\left[4\mathtt{s}(\mathcal{N} - 1) - 1\right]{{\sin }^2}\Theta }\right\rbrace=0,
\end{equation}
which implies that
\begin{equation}
\sin\Theta_{\max}={\sqrt {\frac{{2\mathtt{s}(\mathcal{N} - 1)}}{{4\mathtt{s}(\mathcal{N}-1)-1}}} }.
\end{equation}
Hence, the corresponding maximum evolution speed has the form
\begin{equation}\label{t}
{\mathcal{V}_{\max}} = \mathtt{J}{\mathtt{s}^2}(\mathcal{N} - 1)\sqrt {\frac{{2\mathcal{N}(\mathcal{N} - 1)}}{{4\mathtt{s}(\mathcal{N} - 1) - 1}}}.
\end{equation}
Having established the maximal value of the evolution speed that can be achieved by the $\mathcal{N}$ spin-$\mathtt{s}$ system during evolution, we must also to derive the geodesic distance measured by the squared line element \eqref{o} between the starting state \eqref{d} and the ending state \eqref{e}. To do this, employing equation \eqref{n}, we find
\begin{equation}\label{s}
{\mathtt{\bf{S}}} = \mathtt{s}\sqrt{\frac{{{\xi^2}\mathcal{N}(\mathcal{N} - 1){{\sin }^2}\Theta \left\lbrace {1 + \left[4\mathtt{s}(\mathcal{N}-1) - 1\right]{{\cos }^2}\Theta } \right\rbrace}}{2}}.
\end{equation}
Since the evolution speed \eqref{p} is independent of time, the geodesic distance \eqref{s} varies linearly with the time $t$. As we can see that for $\Theta = 0$ or $\Theta =\pi$, the geodesic distance \eqref{s} cancels $(\mathtt{\bf{S}}=0)$. This is because the relevant state space has a singularity at these points. Furthermore, for ${\Theta=\raise0.7ex\hbox{$\pi$} \!\mathord{\left/
 {\vphantom {\pi 2}}\right.\kern-\nulldelimiterspace}
\!\lower0.7ex\hbox{$2$}}$, the geodesic distance has a local minimum given by 
\begin{equation}\label{v}
 {\mathtt{\bf{S}}}_{\min }= \mathtt{s}\sqrt {\frac{{{\xi ^2}\mathcal{N}(\mathcal{N}-1)}}{2}}.
\end{equation}
Therefore, the smallest possible time (i.e., the optimal time) needed to achieve the time-optimal evolution of the $\mathcal{N}$ spin-$\mathtt{s}$ system is obtained as
\begin{equation}\label{w}
\boldsymbol{\tau} =\frac{{\mathtt{\textbf{S}}}_{\min }}{{\mathcal{V}_{\max}}}= \frac{1}{{2\mathtt{J}\mathtt{s}(\mathcal{N} - 1)}}\sqrt {{\xi ^2}\left[ 4\mathtt{s}(\mathcal{N}-1)-1 \right]},
\end{equation}
which depends only on the maximum speed and minimum distance obtained during an evolutionary process, and not on the parameter $\Theta$. The condition \eqref{w} ensures the time-optimal evolution of the system over the state circle \eqref{o}. In other words, such an evolution is identified by the maximal speed \eqref{t} and the minimal distance \eqref{v} between the points representing the departure and arrival quantum states. Consequently, one can generate the optimal evolution states via the following unitary transformation:
\begin{equation}
\left| {{\Psi _i}} \right\rangle  \to \left| {\Psi \left( {{\bf{\boldsymbol{\tau} }}} \right)} \right\rangle  = {e^{ - i\mathrm{H}{\bf{\boldsymbol{\tau} }}}}\left| {{\Psi _i}} \right\rangle.
\end{equation}
Further, using equation \eqref{w}, the smallest possible time $\boldsymbol{\tau}$ can be expressed in relation to the ordinary time $t$ as
\begin{equation}\label{x}
\boldsymbol{\tau} = \frac{1}{{2\mathtt{s}(\mathcal{N} - 1)}}\sqrt {\left[ 4\mathtt{s}(\mathcal{N} - 1) - 1\right]} t.
\end{equation}
It is worth noting that the ordinary time $t$ corresponds to the circular trajectory defined by the Fubini-Study metric \eqref{o}. From the equation \eqref{x}, we note that the optimal and ordinary times are proportional. More precisely, the increase of the ordinary time is accompanied by the increase of the optimal time. In the case of  $\mathcal{N}=2$ and  $\mathtt{s}={\raise0.7ex\hbox{$1$} \!\mathord{\left/
 {\vphantom {1 2}}\right.\kern-\nulldelimiterspace} 
\!\lower0.7ex\hbox{$2$}}$ (i.e., two-qubit system), the optimal and ordinary times coincide ($\boldsymbol{\tau}=t$), whereas in the case of $\mathcal{N} \ge2 $ and $\mathtt{s}>{\raise0.7ex\hbox{$1$} \!\mathord{\left/
 {\vphantom {1 2}}\right.\kern-\nulldelimiterspace} 
\!\lower0.7ex\hbox{$2$}}$ (i.e., $\mathcal{N}$ qudits system), the optimal time is strictly less than the ordinary time ($\boldsymbol{\tau}<t$). As we can see in the thermodynamic limit $(\mathcal{N} \to \infty)$, the optimal time tends to zero ($\boldsymbol{\tau}\to 0)$. In this picture, the state circle \eqref{o} becomes a straight line, because the radius of the state circle is endless. The same results are obtained when the spin value tends to infinity ($\mathtt{s}\to\infty$). \par
It is also interesting to discuss the solution of the quantum brachistochrone problem which can be obtained by minimizing the time evolution with respect to the parameter $\Theta$. In this instance, one finds
$$\boldsymbol{\tau} =\frac{{\mathtt{\textbf{S}}}_{\min }}{{\mathcal{V}_{\max}}}=\frac{{\mathtt{\textbf{S}}}(\Theta=\pi/2)}{{\mathcal{V}(\Theta=\pi/2)}}=t.$$
This is a trivial case in which the ordinary time coincides with the optimal time $(t=\boldsymbol{\tau})$, which is only valid for a two-qubit system ($\mathcal{N}=2$, $\mathtt{s}=1/2$) and not for general values of $\mathcal{N}$ and $\mathtt{s}$. Furthermore, we could never achieve optimal-time evolution, which is why this approach is inadequate. To end up this section,  we note that it very important to explore the dynamical and geometrical aspects of quantum entanglement by restricting our study to a two spin-$\mathtt{s}$ system under the Ising interaction.

\section{Geometrical and dynamical aspects of the entanglement for two spin-$\mathtt{s}$ system ($\mathcal{N}$=2)}\label{Sec5}
As already mentioned, our goal is to explore the connections between the geometrical features and the amount of quantum entanglement exhibited in our theoretical model. By employing the I-concurrence as a reliable quantifier of quantum entanglement \cite{Rungta2001,Roy2022,Khoury2014}, we will study the entanglement between the two spin-$\mathtt{s}$ system (two-qudit system) through two approaches: The first approach is of geometrical nature which concerns the study of the entanglement effect on geometrical structures established above, namely, the state space geometry, G-curvature, and the geometrical phase.  Second, the dynamic approach deals with the effect of entanglement on the dynamics of the system. Furthermore, we try to solve the quantum brachistochrone problem by means of the degree of quantum entanglement. 
\subsection{Entanglement between two spin-$\mathtt{s}$}
The evolved state of the entire quantum system \eqref{e} can be reduced for a two spin-$\mathtt{s}$ system as
 \begin{align}\label{af}
&\left| {\Psi (t)} \right\rangle  = {(1 + \mathtt{Z}{\mathtt{Z}^*})^{ - 2\mathtt{s}}}\notag\\&\sum\limits_{{\mathsf{m}_1},{\mathsf{m}_2} =  - \mathtt{s}}^\mathtt{s} {{e^{ - i2\xi {\mathsf{m}_1}{\mathsf{m}_2}}}{\mathtt{Z}^{2\mathtt{s} + {\mathsf{m}_1} + {\mathsf{m}_2}}}\sqrt {C_{2\mathtt{s}}^{\mathtt{s} + {\mathsf{m}_1}}C_{2\mathtt{s}}^{\mathtt{s} + {\mathsf{m}_2}}} } \left| {{\mathsf{m}_1},{\mathsf{m}_2}} \right\rangle.
\end{align}
The I-concurrence of a two-qudit state is given by \cite{Oxman2011,Rungta2001}
\begin{equation}\label{ah}
{\mathcal{C}}=\sqrt{2\left(1-\operatorname{Tr}[\rho^{2}_k]\right)},
\end{equation}
where $\rho_k$ is the partial density matrix with respect to spin-$\mathtt{s}$ $k$ ($k=1,2$). To compute the I-concurrence, we start by evaluating the reduced density matrix $\rho_1$ associated to the first spin-$\mathtt{s}$. The density matrix $\rho$ of the entire system \eqref{af} reads as
\begin{equation}
\begin{array}{ll}
\rho=& {(1 + \mathtt{Z}{\mathtt{Z}^*})^{ - 4\mathtt{s}}}\sum\limits_{{\mathsf{m}_1},{\mathsf{m}_2} =  - \mathtt{s}}^\mathtt{s}\sum\limits_{{\mathsf{m}_1^\prime},{\mathsf{m}_2^\prime} =  - \mathtt{s}}^\mathtt{s} {e^{ - i2\xi {\mathsf{m}_1}{\mathsf{m}_2}}}{\mathtt{Z}^{2\mathtt{s} + {\mathsf{m}_1} + {\mathsf{m}_2}}}\\[20px]&{e^{  i2\xi {\mathsf{m}_1^\prime}{\mathsf{m}_2^\prime}}}{\mathtt{Z^*}^{2\mathtt{s} + {\mathsf{m}_1^\prime} + {\mathsf{m}_2^\prime}}} \sqrt {C_{2\mathtt{s}}^{\mathtt{s} + {\mathsf{m}_1}}C_{2\mathtt{s}}^{\mathtt{s} + {\mathsf{m}_2}} C_{2\mathtt{s}}^{\mathtt{s} + {\mathsf{m}_1^\prime}}C_{2\mathtt{s}}^{\mathtt{s} + {\mathsf{m}_2^\prime}} }\;\\[20px]&  \left| {{\mathsf{m}_1},{\mathsf{m}_2}} \right\rangle\left\langle {\mathsf{m}_1^\prime},{\mathsf{m}_2^\prime}\right|.
\end{array}
\end{equation}
It follows that the reduced density matrix $\rho_1$ is given by
\begin{equation}\label{expande}
\begin{array}{ll}
\rho_1=& {\left(1 + \tan^2 \frac{\Theta }{2} \right)^{ - 4\mathtt{s}}}\sum\limits_{{\mathsf{m}_1},{\mathsf{m}_1^\prime} =  - \mathtt{s}}^\mathtt{s}\left[\sum\limits_{{\mathsf{m}_2} =  - \mathtt{s}}^\mathtt{s} {e^{ - i2\xi {\mathsf{m}_1}{\mathsf{m}_2}}}{e^{  i2\xi {\mathsf{m}_1^\prime}{\mathsf{m}_2}}}\right.\\[18px]&\left.{\left(\tan \frac{\Theta }{2}{e^{ - i\Phi }}\right)^{2\mathtt{s} + {\mathsf{m}_1} + {\mathsf{m}_2}}}{\left(\tan \frac{\Theta }{2}{e^{  i\Phi }}\right)^{2\mathtt{s} + {\mathsf{m}_1^\prime} + {\mathsf{m}_2}}} C_{2\mathtt{s}}^{\mathtt{s} + {\mathsf{m}_2}}\right.\\[18px]&\left. \times\sqrt {C_{2\mathtt{s}}^{\mathtt{s} + {\mathsf{m}_1}} C_{2\mathtt{s}}^{\mathtt{s} + {\mathsf{m}_1^\prime}} } \right] \left| {{\mathsf{m}_1}} \right\rangle\left\langle {\mathsf{m}_1^\prime}\right|.
\end{array}
\end{equation}
Adopting the approximation in which we consider the system evolves over a short period of time. In this framework, one can expand the two exponential factors in equation \eqref{expande} up to the second order in $\xi$, we have then
\begin{align}\label{vn}
&{e^{  -i2\xi {\mathsf{m}_1}{\mathsf{m}_2}}}
\approx 1 -2i\xi\mathsf{m}_1\mathsf{m}_2-2\xi^2(\mathsf{m}_1\mathsf{m}_2)^2,\notag\\&{e^{  i2\xi {\mathsf{m}_1^\prime}{\mathsf{m}_2}}}\approx 1+ 2i\xi\mathsf{m}_1^\prime\mathsf{m}_2-2\xi^2(\mathsf{m}_1^\prime\mathsf{m}_2)^2.
\end{align}
Putting equations \eqref{vn} into \eqref{expande}, the reduced density matrix $\rho_1$ becomes
\begin{widetext}
\begin{equation}\label{expandye}
\begin{array}{ll}
\rho_1={\left(1+\tan^2 \frac{\Theta}{2} \right)^{- 4\mathtt{s}}}&\sum\limits_{{\mathsf{m}_{1}},{\mathsf{m}_{1}^\prime} =-\mathtt{s}}^\mathtt{s}\left[\sum\limits_{{\mathsf{m}_2}=- \mathtt{s}}^\mathtt{s} \left[ 1+2i\xi(\mathsf{m}_1^\prime-\mathsf{m}_1)\mathsf{m}_{2}+4\xi^2\mathsf{m}_1\mathsf{m}_1^\prime \mathsf{m}_2^2-2\xi^2(\mathsf{m}_1^2+{\mathsf{m}_1^\prime}^2)\mathsf{m}_2^2\right] {\left(\tan^2 \frac{\Theta }{2}\right)^{\mathtt{s}  + {\mathsf{m}_2}}}C_{2\mathtt{s}}^{\mathtt{s} + {\mathsf{m}_2}}\right.\\[18px]&\left.{\left(\tan \frac{\Theta }{2}{e^{ - i\Phi }}\right)^{\mathtt{s} + {\mathsf{m}_1} }}{\left(\tan \frac{\Theta }{2}{e^{  i\Phi }}\right)^{\mathtt{s} + {\mathsf{m}_1^\prime}}}\sqrt {C_{2\mathtt{s}}^{\mathtt{s} + {\mathsf{m}_1}} C_{2\mathtt{s}}^{\mathtt{s} + {\mathsf{m}_1^\prime}} } \right]\left| {{\mathsf{m}_1}} \right\rangle\left\langle {\mathsf{m}_1^\prime}\right|.
\end{array}
\end{equation}
Using the binomial theorem \eqref{ppm} and its derivatives
\begin{equation}
\sum\limits_{{\mathsf{m}_k} =  - \mathtt{s}}^\mathtt{s} \mathsf{m}_k{{{\left( {{{\tan }^2}\frac{\Theta }{2}} \right)}^{\mathtt{s} + {\mathsf{m}_k}}}} C_{2\mathtt{s}}^{\mathtt{s} + {\mathsf{m}_k}} = -\mathtt{s}{\left( {1 + {{\tan }^2}\frac{\Theta }{2}} \right)^{2\mathtt{s}}}\cos\Theta,
\end{equation}
and
\begin{equation}
\begin{aligned}\sum\limits_{{\mathsf{m}_k} =  - \mathtt{s}}^\mathtt{s} \mathsf{m}_2^2{{{\left( {{{\tan }^2}\frac{\Theta }{2}} \right)}^{\mathtt{s} + {\mathsf{m}_k}}}} C_{2\mathtt{s}}^{\mathtt{s} + {\mathsf{m}_k}} = \left( \mathtt{s}^2-\dfrac{\mathtt{s}}{2}(2\mathtt{s}-1)\sin^2\Theta  \right)\times{\left( {1 + {{\tan }^2}\frac{\Theta }{2}} \right)^{2\mathtt{s}}},
\end{aligned}
\end{equation}
we obtain the reduced density matrix \eqref{expandye} of the form
\begin{equation}\label{expandyle}
\begin{array}{ll}
\rho_1= {\left(1 + \tan^2 \frac{\Theta }{2} \right)^{ - 2\mathtt{s}}}&\sum\limits_{{\mathsf{m}_1},{\mathsf{m}_1^\prime} =  - \mathtt{s}}^\mathtt{s}\left[ 1-2i\mathtt{s}\xi(\mathsf{m}_1^\prime-\mathsf{m}_1)\cos\Theta -2\xi^2(\mathsf{m}_1-\mathsf{m}_1^\prime)^2\left( \mathtt{s}^2-\dfrac{\mathtt{s}}{2}(2\mathtt{s}-1)\sin^2\Theta  \right)\right]{\left(\tan \frac{\Theta }{2}{e^{ - i\Phi }}\right)^{\mathtt{s} + {\mathsf{m}_1} }}\\[18px]& \times{\left(\tan \frac{\Theta }{2}{e^{  i\Phi }}\right)^{\mathtt{s} + {\mathsf{m}_1^\prime}}} \sqrt {C_{2\mathtt{s}}^{\mathtt{s} + {\mathsf{m}_1}} C_{2\mathtt{s}}^{\mathtt{s} + {\mathsf{m}_1^\prime}} }  \left| {{\mathsf{m}_1}} \right\rangle\left\langle {\mathsf{m}_1^\prime}\right|.
\end{array}
\end{equation}
\end{widetext}
Inserting equation \eqref{expandyle} into \eqref{ah} and employing the binomial theorem and its derivatives, we obtain easily the I-concurrence of the two spin-$\mathtt{s}$ system as follows
\begin{equation}\label{ak}
{\mathcal{C}}= 2\xi \mathtt{s}\sin^2\Theta.
\end{equation}
Thus, we find that the entanglement between the two spins is affected by the dynamical parameters $(\Theta, \xi)$. This means that the entanglement evolution depends on the path followed by the system over the relevant state space. Otherwise, each point of the state manifold corresponds to a particular degree of entanglement defined by these parameters. Moreover, taking into account the approximation used in equation \eqref{vn}, we can see that for $\Theta=
{\raise0.7ex\hbox{$\pi $} \!\mathord{\left/
 {\vphantom {\pi  2}}\right.\kern-\nulldelimiterspace}
\!\lower0.7ex\hbox{$2$}}$ and $\xi=\xi_{\max}^\prime=\mathtt{J}dt$ (i.e., $\xi\in[0,\xi_{\max}^\prime] \;\text{where we take the exchange constant}\; 0<\mathtt{J}\le 1$ for the sake of precision), the I-concurrence ${\mathcal{C}}$ reaches the maximum value $({\mathcal{C}}_{\max}=2\mathtt{s}\xi_{\max}^\prime)$, which, in turn, depends on the spin value. For $\Theta=0$ or $\Theta=\pi$, we can never achieve an entangled state. This happens because the initial state $\left| {{\Psi _i}} \right\rangle  = \left| { \mathtt{s}, \mathtt{s}} \right\rangle$ is an eigenstate of the two spin-$\mathtt{s}$ system. Geometrically, we can say that the G-curvature \eqref{h} presents a singularity in these two points. However, for all other points (e.g., $\Theta\in]0,\pi[$), the entanglement is well defined and evolves linearly with time for a predefined value of $\Theta$. 
\subsection{Geometrical picture of the entanglement}
To highlight the geometrical appearance of the quantum correlations between the two spins under consideration, we propose a detailed study visualizing the interplay between the entanglement and the geometrical features established above. Using equations \eqref{ak} and \eqref{m}, the metric tensor, defining the space of two-spin-$\mathtt{s}$ states, can be expressed in terms of the I-concurrence as
\begin{widetext}
\begin{align}\label{al}
d{{\mathtt{\bf{S}}}}^2& = \frac{\mathtt{s}}{{2{\xi ^2}\mathcal{C}(2\mathtt{s}\xi  -\mathcal{C})}}\left[ \frac{\xi ^2}{2}d{\mathcal{C}^2} - \xi \mathcal{C}d\mathcal{C}d\xi  + \left( \frac{\mathcal{C}^2}{2} + { \xi_{\max}^\prime}\mathcal{C}(2\mathtt{s}\xi  -\mathcal{C})\left(1 + \left(4\mathtt{s} - 1\right)\left(1 - \frac{{ {\xi _{\max}^\prime}\mathcal{C}}}{{2\mathtt{s}{\xi ^2}}} \right)\right) \right)d{\xi ^2} \right].
\end{align}
\end{widetext}
Therefore, we managed to parametrize the two-spin state space in relation to the entanglement degree as well as the evolution time, being two quantifiable physical quantities. This shows the ability to investigate experimentally some geometrical and dynamical features identifying the state space \eqref{al}, such as the state space geometry, the geometrical phase, the evolution speed, and the geodesic distance between the entangled states. On the other hand, the entanglement provides the possibility to lower the state space dimensions, for example, the two spin-$\mathtt{s}$ states of the same entanglement degree constitute a curved one-dimensional manifold identified by the following metric tensor:
\begin{align}\label{as}
d{{\mathtt{\bf{S}}}}^2 = \frac{\mathtt{s}}{{2{\xi ^2}\mathcal{C}(2\mathtt{s}\xi  -\mathcal{C})}}&\left[ \frac{\mathcal{C}^2}{2} + { \xi_{\max}^\prime}\mathcal{C}(2\mathtt{s}\xi  -\mathcal{C})\times\right.\notag\\&\left.\left(1 + (4\mathtt{s} - 1)\left( 1 - \frac{{ {\xi _{\max}^\prime}\mathcal{C}}}{{2\mathtt{s}{\xi ^2}}} \right)\right) \right]d{\xi ^2}.
\end{align}
Hence the relevance of quantum correlations for the adaptation of the state space geometry. In this respect, we can also explore the entanglement effect on the G-curvature of two-spin state space \eqref{al}. Indeed, introducing the equation \eqref{ak} into \eqref{ay}, we give the G-curvature in terms of the I-concurrence as follows
\begin{equation}\label{ar}
K = \frac{2}{\mathtt{s}}\left[ {2 - \frac{{(4\mathtt{s} - 1)\left( {1 - \tilde \xi \frac{{\mathcal{C}}}{{{{\mathcal{C}}_{\max }}}}} \right) + 2\mathtt{s} + 1}}{{{{\left[ {(4\mathtt{s} - 1)\left(1 - \tilde \xi \frac{{\mathcal{C}}}{{{{\mathcal{C}}_{\max }}}}\right) + 1} \right]}^2}}}} \right],
\end{equation}
where $\tilde{\xi}={{{ \xi_{\max}^\prime}} \mathord{\left/
 {\vphantom {{{ \xi_{\max}^\prime}} \xi }} \right.
 \kern-\nulldelimiterspace} \xi }$. This result proves once again the sensitivity of the state space geometry against the quantum correlations existing between the two spins. 
 \begin{figure}[h]
\begin{center}
\includegraphics[scale=0.52]{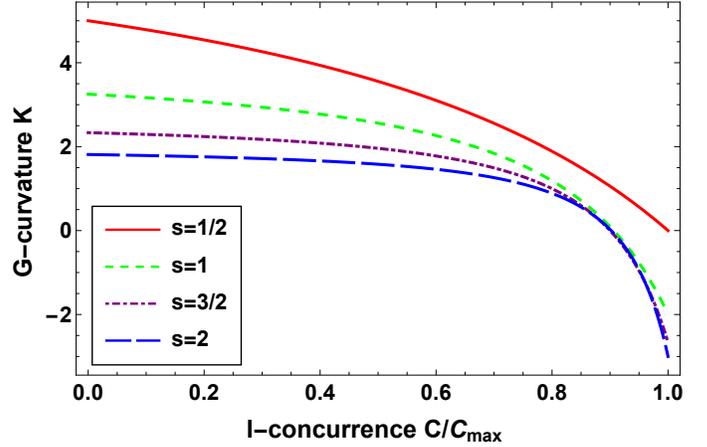}
\caption{The G-curvature \eqref{ar} versus the I-concurrence \eqref{ak}  for some spin values with $\tilde{\xi}=1$. }\label{am}
\end{center}
\end{figure}

To get an insight into the G-curvature with respect to quantum entanglement, we plotted the behavior of \eqref{ar} versus the I-concurrence for several spin values and fixed $\tilde{\xi}=1$ in Fig.\eqref{am}. We observe that increasing the degree of entanglement between the two spins leads to a decrease in the state space curvature. Further, for $\mathtt{s} >1/2$, the G-curvature takes negative values. This can be interpreted by the fact that the increase of the quantum correlations between the two spins compactifies the corresponding state space \eqref{as}. As we can see that the disentangled states ($\mathcal{C}=0$) are located in the maximal curvature areas
\begin{equation}
{K_{\max}} = \frac{2}{\mathtt{s}}\left[ {2 - \frac{3}{{8\mathtt{s}}}} \right],
\end{equation}
whereas the maximally entangled states $(\mathcal{C}=\mathcal{C}_{\max})$ are located in the minimal curvature areas
\begin{equation}
{\textit{K}_{\min }} = \frac{2}{\mathtt{s}}\left[ {2 - \frac{{(4\mathtt{s} - 1)( {1 - \tilde \xi }) + 2\mathtt{s} + 1}}{{{{\left[ {(4\mathtt{s} - 1)(1 - \tilde \xi ) + 1} \right]}^2}}}} \right].
\end{equation}
In this way, we conclude that the amount of entanglement of the two spin-$\mathtt{s}$ system identifies the physical states of the system in the quantum phase space \eqref{al}. This allows us to obtain information about the geometry of state manifold through the entangled states. Besides, we see that for high spin values $(\mathtt{s} \to \infty$), the G-curvature vanishes $(\textit{K}=0)$ at any point of the state space, hence this space becomes flat. In the same framework, we discuss the interplay between the geometrical phase and the entanglement. Indeed, putting equation \eqref{ak} into \eqref{an}, the geometrical phase acquired by the two spin-$\mathtt{s}$ system can be expressed in terms of the I-concurrence as
\begin{small}
\begin{align}\label{ao}
	&{\boldsymbol{\Phi}_\text{g}}=2\xi {\mathtt{s}^2}\left( {1 - \tilde \xi \frac{{\mathcal{C}}}{{{{\mathcal{C}}_{\max}}}}} \right)\notag\\&-\arctan \left( {\frac{{4\xi {\mathtt{s}^2}\left( {1 - \tilde \xi \frac{{\mathcal{C}}}{{{{\mathcal{C}}_{\max }}}}} \right)}}{{2 - {\xi ^2}{\mathtt{s}^2}(2\mathtt{s} -1 )\left( { {(2\mathtt{s} - 1)}{{\tilde \xi }^2}\frac{{{{\mathcal{C}}^2}}}{{{\mathcal{C}}_{\max }^2}} - 4\mathtt{s}\tilde \xi \frac{{\mathcal{C}}}{{{{\mathcal{C}}_{\max }}}} + 4{\mathtt{s}^2}} \right)}}} \right).
\end{align}
\end{small}
\begin{figure}[h]
\begin{center}
\includegraphics[scale=0.52]{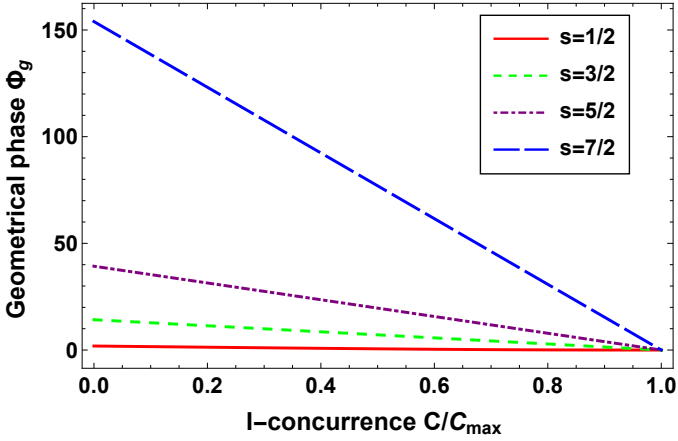}
\caption{The geometrical phase \eqref{ao} versus the I-concurrence \eqref{ak} for some spin values with $\tilde{\xi}=1$. }\label{ap}
\end{center}
\end{figure}

In Fig.\eqref{ap}, we depict the geometrical phase behavior of the two spin-$\mathtt{s}$ system versus the quantum entanglement for various spin values with $\tilde{\xi}=1$. From the results reported in this figure, it can be clearly seen that the geometrical phase varies as a decreasing affine function with increasing entanglement degree. This behavior can be explained by returning to the expression of the geometrical phase \eqref{ao}, where the contribution of the dynamical phase is clearly more dominant than that of the global phase. Notice that the disentangled states constitute the category of the two-spin states  most accessible to accumulate the geometric phase maximum, while the maximally entangled states form the category of states that never accumulate it. The intermediate entangled states are situated between these two types of categories. Accordingly, we find that the existence of quantum correlations favors the loss of the geometrical phase during the evolution on the state space \eqref{al}. This result agrees with the result demonstrated above that quantum correlations decrease the G-curvature \eqref{ar}, because the geometrical phase is a consequence of such curvature. On the other hand, we notice that the larger the spin value, the greater the gain (or the loss) of the geometric phase. Therefore, we conclude that the entanglement and the spin value are two interesting keys for controlling the gain or loss of the Berry phase of the two spin-$\mathtt{s}$ system.

 \subsection{Dynamical picture of the entanglement}
\begin{figure}[h]
\begin{center}
\includegraphics[scale=0.52]{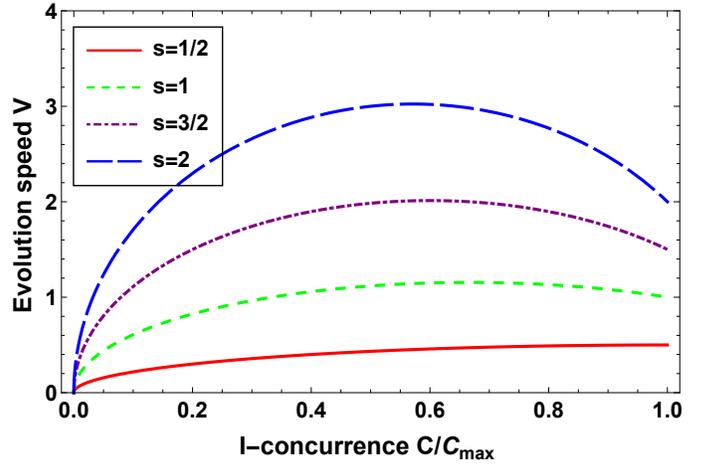}
\caption{The evolution speed \eqref{at} versus the I-concurrence \eqref{ak} for some spin values with  $\tilde{\xi}=1$ and $\mathtt{J}=1$. }\label{akk}
\end{center}
\end{figure}

At the end of this section, we will investigate the interplay between the entanglement of the two spin-$\mathtt{s}$ system and the corresponding dynamics over the quantum phase space \eqref{al}. As a result, we attempt to solve the quantum brachistochrone problem based on the degree of quantum entanglement between the two interacting spins. For this purpose, by inserting the equation \eqref{ak} into \eqref{n} for two-spin system $(\mathcal{N}=2)$, the evolution speed can be rewritten in terms of the I-concurrence as
 \begin{equation}\label{at}
\mathcal{V} = \mathtt{J}\mathtt{s}\sqrt {\tilde \xi \frac{{\mathcal{C}}}{{{{\mathcal{C}}_{\max }}}}\left( {4\mathtt{s} - (4\mathtt{s} - 1)\tilde \xi \frac{{\mathcal{C}}}{{{{\mathcal{C}}_{\max }}}}} \right)}.
\end{equation}
Thereby, we succeeded in connecting the evolution speed of two spin-$\mathtt{s}$ system with the degree of entanglement. Otherwise, the expression \eqref{at} translates the connection the unitary evolution of the system and the dynamics of quantum correlations. Therefore, we deduce that the information about the dynamics of the system can be obtained through its entangled states. The evolution speed behavior with respect to the I-concurrence is displayed in Fig.\eqref{akk}.

 We note that for $\mathtt{s}>1/2$ the variation of the evolution speed is divided into two parts: the first part displays the evolution speed increase of the two spin-$\mathtt{s}$ system until it reaches its maximum value $\mathcal{V}_{\max}={{2\mathtt{J}{\mathtt{s}^2}} \mathord{\left/
 {\vphantom {{2J{\mathtt{s}^2}} {\sqrt {4\mathtt{s} - 1} }}} \right.
 \kern-\nulldelimiterspace} {\sqrt {4\mathtt{s} - 1} }}$
that matches the entanglement degree $\mathcal{C}=\mathcal{C}^\prime={{2\mathtt{s}{\mathcal{C}}_{\max }} \mathord{\left/
 {\vphantom {{2\mathtt{s}{\mathcal{C}}_{\max }} {(4\mathtt{s} - 1)\tilde \xi }}} \right.
 \kern-\nulldelimiterspace} {(4\mathtt{s} - 1)\tilde \xi }}$. This means that the quantum correlations, in this first part, accelerate the evolution of the system over the quantum state space \eqref{al}. The second part concerns the interval $\mathcal{C}\in[\mathcal{C}^\prime,{\mathcal{C}}_{\max }]$ where the speed has reversed its variation, it decreases continuously until it reaches its local minimum $\mathcal{V}({\mathcal{C}}={\mathcal{C}}_{\max })$. This signifies that the quantum correlations, in this second part, decelerate the evolution of the system. Hence, the dynamics of the system undergoes a phase change at the critical point $\mathcal{C}=\mathcal{C}^\prime$. The special case $\mathtt{s}=1/2$ includes only the first part because $\mathcal{C}^\prime= {\mathcal{C}}_{\max }$. The speed \eqref{at} is also affected by the spin value. Indeed, we notice that the larger the spin value, the faster the two spin-$\mathtt{s}$ system evolves. Accordingly, we conclude that the entanglement and the spin value are two essential physical parameters for controlling the evolution speed of two interacting spins under the Ising model. Employing equation \eqref{n}, we derive the geodesic distance covered by the two spin-$\mathtt{s}$ system in terms of the I-concurrence, it is given by
\begin{equation} \label{au}
{\textbf{S}}= \mathtt{s}\sqrt {{\xi_{\max}^\prime} \xi \frac{{{\mathcal{C}}}}{{{{{\mathcal{C}}}_{\max }}}}\left( {4\mathtt{s} - (4\mathtt{s} - 1)\tilde \xi \frac{{{\mathcal{C}}}}{{{{{\mathcal{C}}}_{\max }}}}} \right)}.
\end{equation}
Therefore, we were able to relate the distance measured by the Fubini-Study metric \eqref{al} with the entanglement. This provides the possibility to measure  experimentally the distance between the entangled states (or the evolution speed) on the quantum state manifold \eqref{al}.
\begin{figure}[h]
\begin{center}
\includegraphics[scale=0.52]{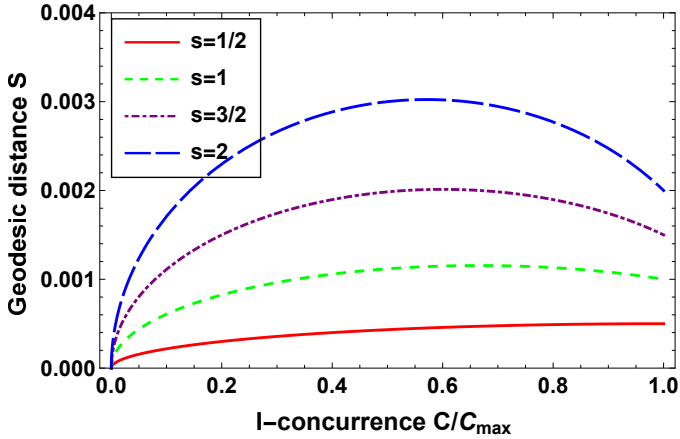}
\caption{The geodesic distance \eqref{au} versus the I-concurrence \eqref{ak} for some spin values with  $\tilde{\xi}=1$, $\mathtt{J}=1$ and $\xi_{\max}^\prime=10^{-3}$. }\label{akp}
\end{center}
\end{figure}

 Comparing the Figs.\eqref{akk} and \eqref{akp}, we conclude that the geodesic distance \eqref{au} has, with respect to the entanglement, the same behavior as the evolution speed \eqref{at} and therefore the same conclusions can be obtained. Let us now resolve the quantum brachistochrone problem which is related to finding the optimal-time evolution for the two spin-$\mathtt{s}$ system. For this purpose, using the equations \eqref{at} and \eqref{au}, the smallest possible duration needed to realize the time-optimal evolution is given in relation with the I-concurrence by
 \begin{small}
 \begin{equation}\label{aulll}
\boldsymbol{\tau}_{\mathcal{C}}  = \frac{{{\textbf{S}}}}{{{\mathcal{V}_{\max }}}} = \frac{1}{{2J\mathtt{s}}}\sqrt {{\xi_{\max}^\prime} \xi \frac{{\mathcal{C}}}{{{{\mathcal{C}}_{\max }}}}(4\mathtt{s} - 1)\left( {4\mathtt{s} - (4\mathtt{s} - 1)\tilde \xi \frac{{\mathcal{C}}}{{{{\mathcal{C}}_{\max }}}}} \right)}.
\end{equation}
 \end{small}
 From this last expression \eqref{au}, we see that for ${\mathcal{C}}=0$ the optimal time vanishes ($\boldsymbol{\tau}=0$), this is because the evolved state \eqref{af} coincides with the separable initial state $\left| {{\Psi _i}} \right\rangle  = {(1 + \mathtt{Z}{\mathtt{Z}^*})^{ - 2\mathtt{s}}}\left\| {\left. {{\mathtt{Z}_1},{\mathtt{Z}_2}} \right\rangle } \right.$ (i.e., no evolution). For the critical entanglement degree ${\mathcal{C}} = {\mathcal{C}}^\prime$ the optimal time reaches its maximum value $(\boldsymbol{\tau}=t)$, meaning that the optimal evolution of two spin-$\mathtt{s}$ system coincides with the ordinary evolution, while for $\mathcal{C}\in\left]0,{\mathcal{C}}^\prime\right[\; \cup\; ]{\mathcal{C}}^\prime,{\mathcal{C}}_{\max}[$ the optimal time is strictly inferior to the optimal time $(\boldsymbol{\tau}_{\mathcal{C}}<t)$.
\begin{figure}[h]
\begin{center}
\includegraphics[scale=0.52]{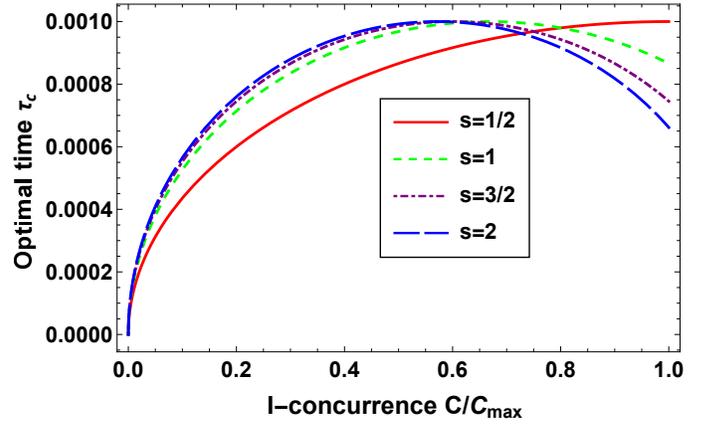}
\caption{The optimal time \eqref{aulll} versus the I-concurrence \eqref{ak} for some spin values with  $\tilde{\xi}=1$, $\mathtt{J}=1$ and $\xi_{\max}^\prime=10^{-3}$.}\label{akf}
\end{center}
\end{figure}

The behavior of the optimal time \eqref{aulll} with respect to the entanglement is depicted in Fig.\eqref{akf}, we note that the smaller the entanglement degree (or the spin value $\mathtt{s}$) of the two spins $(\mathcal{C}\to 0)$, the smaller the optimal time $(\boldsymbol{\tau}_{\mathcal{C}} \to 0)$. 
As a result, we conclude that entanglement and spin value are two intriguing physical quantities in the realization of the optimal-time evolution for two interacting spins under the Ising model. These evolutions are extremely important in the field of quantum computation, particularly in the construction of quantum circuits required to implement quantum gates \cite{Nielsen2006,DiVincenzo1998,Vartiainen2004}.
\section{Conclusion and outlook}\label{sec6}
To summarize, we studied a physical system of $\mathcal{N}$ interacting spin-$\mathtt{s}$ under long-range Ising model. We have presumed that the departure state is the tensor product of $\mathcal{N}$ SU(2) coherent states. After applying the unitary evolution operator, the resulting evolved state is defined by three degrees of freedom, namely, the spherical angles $(\Theta,\Phi)$ and the time $t$. We give analytically the Fubini-Study metric and found that the state space is a curved two-dimensional manifold. Additionally, using the Gauss-Bonnet theorem we proved that this space is of spherical topology. We computed the geometrical phase accumulated by the $\mathcal{N}$ spin-$\mathtt{s}$ system for arbitrary and cyclic evolution processes. This result is obtained by assessing the difference between the global and dynamical phase. We concluded that in the arbitrary evolution case the geometrical phase evolves non-linearly with time. Geometrically, we discovered that it depends on both the evolutionary path followed by the system as well as the geometry taken by the corresponding quantum phase space. In the cyclic evolution case, the integral of the global phase accumulated by the system along the evolutionary process vanishes, and therefore the geometrical phase coincides with the dynamical phase which evolves linearly with time, hence, the possibility to measure the geometrical phase experimentally. The same result can be obtained in the thermodynamic limit $(\mathcal{N}\to \infty)$. In addition, we examined the evolution speed, geodesic distance between the quantum states, and we resolved the quantum brachistochrone problem for
the $\mathcal{N}$ spin-$\mathtt{s}$ system. We demonstrated the optimal time in terms of ordinary time. As a result, we inferred that for $\mathcal{N}=2$ and $\mathtt{s}=1/2$ (i.e., the two spin-$1/2$ system), the optimal and arbitrary time coincide, whereas for $\mathcal{N}>2$ and $\mathtt{s}>1/2$, the optimal time is strictly inferior to the ordinary time. Thus, the number of particles and the spin value are two essential parameters to realize the optimal-time evolution. Moreover, it is important to see that in the thermodynamic limit $(\mathcal{N}\to \infty)$, the optimal time is close to zero. The same results can be achieved for large spin values.\par 

On the other hand, by narrowing the system to a two interacting spin-$\mathtt{s}$ under the Ising model, we analyzed the entanglement from two perspectives: the first perspective is of geometrical type, in which we established the Fubini-Study metric in relation with the I-concurrence as a quantifier of the quantum correlations. Moreover, we showed that the existence of quantum correlations between the two interacting spins decreases the curvature of the relevant state space. In addition, the entanglement amount identifies the physical states on the state space, for instance, the states of maximum entanglement are located in the minimum curvature regions, while the disentangled states are located in the maximum curvature regions. We also demonstrated that the existence of quantum correlations disfavors the accumulation of the geometrical phase. Thus, we managed to illustrate the interplay between the associated geometrical structures and the quantum entanglement phenomenon. The second perspective is of dynamical type, in which we related the evolution speed with the entanglement, we found that the speed has two different behaviors with respect to the critical entanglement degree $\mathcal{C}^\prime$ $(\mathcal{V}(\mathcal{C}^\prime)=\mathcal{V}_{\max})$: the first one is in the interval [0, $\mathcal{C}^\prime$], where the existence of the quantum correlations accelerates the evolution of the system, while the second one is in the interval $[\mathcal{C}^\prime$, $\mathcal{C}_{\max}]$, where the existence of the quantum correlations decelerates the evolution of the system. The same behavior is observed for the geodesic distance between the quantum states. Finally, we solved the quantum barchistochrone problem using the degree of entanglement between the two spins. We concluded that the entanglement and the spin value are two essential parameters to realize the optimal-time evolution for two interacting spin-$\mathtt{s}$ under Ising model. Thus, we were able to highlight, to a large extent, the relation between the quantum entanglement and the geometrical and dynamical structures identifying the state space of the two spin-$\mathtt{s}$ system under consideration.\par
As an extension of these results, it is extremely important to generalize our study by examining the connection between these structures and the quantum entanglement when the system consists of more than two qudits ($\mathcal{N}>2$). Thereby, it is interesting to examine the connection between these structures and other quantum resources, such as quantum coherence and correlations beyond quantum entanglement using various measures. Some such quantifiers provide analytical expressions for multi-qudit systems, and thus we can find their links with the relevant features. We hope to address these issues in upcoming work.

\end{document}